# A Modified ck-Secure Sum Protocol for Multi-Party Computation

Rashid Sheikh, Beerendra Kumar, Durgesh Kumar Mishra

**Abstract**—Secure Multi-Party Computation (SMC) allows multiple parties to compute some function of their inputs without disclosing the actual inputs to one another. Secure sum computation is an easily understood example and the component of the various SMC solutions. Secure sum computation allows parties to compute the sum of their individual inputs without disclosing the inputs to one another. In this paper, we propose a modified version of our ck-Secure Sum protocol with more security when a group of the computing parties conspire to know the data of some party.

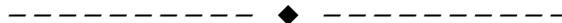

## 1 INTRODUCTION

The huge growth of the Internet and its easy access by common man created opportunities for joint computations by multiple parties. All the participating parties for the sake of their mutual benefit want to compute the common function of their inputs but at the same time they are worried about the privacy of their data. This subject of the information security is called Secure Multi-Party Computation (SMC). This subject has two goals; one is the privacy of the individual data inputs and another is the correctness of the result. Mainly two models exist in the literature for the analysis of the SMC problems. Ideal model of the SMC uses a Trusted Third Party (TTP) apart from the participating parties. Parties supply their inputs to the TTP. Computation of the function is done by the TTP and then the result is sent to all the parties. In this paradigm the trustworthiness of the TTP is critically important because if the TTP turns corrupt, it can supply the private inputs of one party to others. But it is extensively used model of the SMC due to its easy implementation and the protocols available which prevent the TTP to act maliciously. Real model of the SMC does not use any TTP but the parties themselves agree on some protocol for the computation. The party behavior in the SMC is important to consider. An honest party follows the protocol and respects the privacy of other parties. A semi honest party follows the protocol but also tries to learn other information than the result. The corrupt party neither follows the protocol nor respects the privacy of other parties. Different protocols are needed for different SMC models and the behavior of the party. Solutions are available for SMC problems using cryptographic techniques, randomization techniques and anonymization methods. The subject of SMC has been evolved from two party comparison problems [1] to multiparty image template matching problems. Many specific SMC problems have been defined and analyzed by researchers like Private Information Retrieval (PIR), Selective Function Evaluation, Privacy-Preserving Database Query, Privacy-Preserving Geometric Computation, Privacy-Preserving Statistical Analysis, Privacy-Preserving Intrusion Detection and Privacy-Preserving Cooperative Scientific Computation. Based on these general SMC problems many real life applications emerged like Privacy-Preserving Electronic Voting, Privacy-Preserving Bidding and Auctions, Privacy-Preserving Social Network Analysis, Privacy-Preserving Signature and Face Detection, etc.

Secure sum computation problem of SMC can be defined as: How multiple parties can compute the sum of their input values without disclosing actual values to one another. Secure sum can work as the tool for the SMC solutions in the privacy preserving distributed data mining problems [2]. Clifton *et al.* proposed a secure sum protocol using random numbers [2]. We proposed novel secure sum protocols with more security in [3, 4, 5]. In this paper, we propose a novel changing neighbors approach for achieving more security in case a group of the parties collude to know the private data of some other party.

## 2 RELATED WORK

The subject of SMC started in 1982 when Yao proposed millionaires' problem in which two millionaires wanted to know who was richer without disclosing individual wealth to one another [1]. The concept was extended by Goldreich *et al.* [6]. They used circuit evaluation protocols for secure computation. Many real life applications of SMC emerged like Private Information Retrieval (PIR) [7, 8], Privacy-preserving data mining [9, 10], Privacy-preserving geometric computation [11], Privacy-preserving scientific computation [12], Privacy-preserving statistical analysis [13] etc. An excellent review of SMC is provided by Du *et al.* in [14] where they developed a

———————————————

- *Rashid Sheikh and Beerendra Kumar are with the Sri Satya Sai Institute of Science and Technology, Sehore, India*
- *Durgesh Kumar Mishra is with the Acropolis Institute of Technology and Research, Indore, India*

framework for problem discovery. A study of SMC problems with a focus on telecommunication systems is provided by Oleshchuk et al. [15]. Anonymity enabled solution was proposed by Mishra et al. [16] where the identities of the parties were hidden for privacy.

In a paper, Clifton et al. [2] proposed a toolkit of components for solution to SMC problems. They pointed out that one of components of the toolkit for SMC is the secure sum computation. Secure sum computation is used in many distributed data mining algorithms where many distributed sites compute sum of values. The secure sum protocol proposed by Clifton et al. [2] used random numbers for privacy of individual data inputs. In this protocol any two parties $P_{i-1}$ and $P_{i+1}$ can collude to know the secret data of $P_i$ by performing only one computation. We proposed *k-Secure Sum* Protocol and *Extended k-Secure Sum* Protocol [4] where the probability of data leakage is significantly reduced by breaking the data block of individual party into number of segments. The probability decreases as the number of segments in a data block is increased. In another paper we proposed a zero probability protocol *ck-Secure Sum* Protocol for two colluding neighbors [3].

In this paper we proposed zero probability protocol for secure sum computation namely *Modified ck-Secure Sum* Protocol which is an extension of the previous protocol. It provides preservation of privacy when more they two parties collude to know the secret data of some party.

## 3 PROPOSED PROTOCOL

In this section, we first describe our *ck-Secure Sum* Protocol [3] and then it's extended version the *Modified ck-Secure Sum* Protocol. In both of these protocols the data block is divided into a fixed number of segments and the computation is performed on these segments taking one segment at a time.

### 3.1 The *ck-Secure Sum* Protocol

In *k-secure sum* protocol [4] a middle party can be hacked by two neighbor parties with some probability. The technique for *ck-Secure Sum* Protocol [3] is that we change the neighbors in each round of segment computation. Thus it is guaranteed that no two semi honest parties can know all the data segments of a victim party. In this protocol each of the parties breaks the data block into $k = n-1$ segments where $n$ is the number of parties involved in secure sum computation. We select $P_1$ as the protocol initiator. The position of the protocol initiator is kept fixed in all the rounds of computation. For the first round of the computation parties are arranged serially as $P_1, P_2, …, P_n$. The protocol initiator starts computation to get the sum of first segments of each party. For this computation our *k-Secure Sum* protocol [4] is used. Now, $P_2$ exchanges its position with $P_3$ and second round of computation is performed. Now, $P_2$ exchanges its position with $P_4$ and so on. Formally, in $i^{th}$ round of the computation $P_2$ exchanges its position with $P_{i+1}$ until $P_n$ is reached. In each round of computation, segments are added and the partial sum is passed to the next party until all the segments are added. Finally, the sum is announced by the protocol initiator party. Snapshots for a four-party case are shown in Fig. 1. The *ck-Secure Sum* Protocol provides privacy against two colluding neighbors. Its analysis shows that when more than two parties collude, they can know the data of some party. The protocol initiator can be attacked by more than two parties that maliciously cooperate to know secret data of the protocol initiator. But for that also a specific combination of the parties must join against the protocol initiator. Any party who moves its position cannot be attacked by any group of the parties.

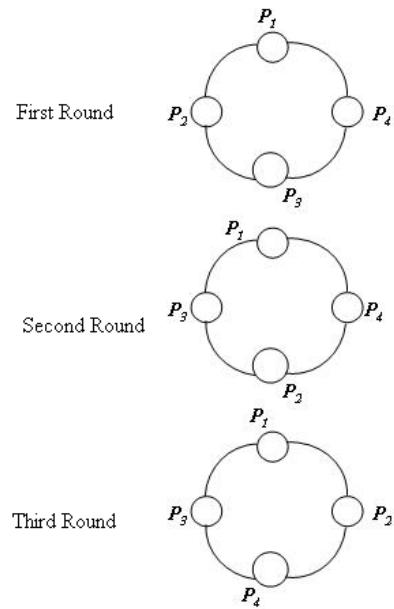

Fig. 1. Snapshots of *ck-Secure Sum Protocol* for four-party case [3].

### 3.2 The *Modified ck-Secure Sum* Protocol

In this protocol mainly two modifications to the *ck-Secure Sum* Protocol are done:

1. The number of segments $k$ is kept equal to the number of the parties $n$.
2. The protocol initiator party moves throught the ring.

### 3.2.1 Informal Description of the *Modified ck-Secure Sum* Protocol

All the parties are arranged in a unidirectional ring. Each party divides its data block into $k$ segments which is equal to the number of the parties. The party $P_1$ is selected as the protocol initiator party for all the rounds of the computation. This party starts computation by sending first data segment to the next party in the ring. The next party adds its segment to the received segment and the sum is passed to the next party. This process continues until all the segments are added. After receiving the sum of the first segments of all the parties, the protocol initiator $P_1$ exchanges its position with $P_2$ and then it sends the sum of its segment and the previous received sum to the next party in the newly arranged ring. At the end of this round, the protocol initiator receives the sum of two segments of all the parties. Now, $P_1$ exchanges its position with $P_3$ and so on until $P_n$ is reached. Finally the sum is announced by the protocol initiator. Fig. 2 depicts how $P_1$ moves within the ring and Fig. 3 depicts architecture for four-party case where each of the parties breaks its data block into four segments. The movement of the parties is just a logical one; it can be implemented by breaking and making the links between the parties.

### 3.2.2 Formal Description of the *Modified ck-Secure Sum* Protocol

The *Modified ck-Secure Sum* Protocol is an extension of *ck-Secure Sum* Protocol [3] and is based on changing neighbors in each round of segment computation. The party $P_1$ is selected as the protocol initiator party which starts the computation by sending the first data segment. The party $P_1$ traverses towards $P_n$ in each round of the computation. The number of parties for this protocol must be four or more. When all the rounds of segment summation are completed the sum is announced by the protocol initiator party.

The algorithm: *Modified ck-Secure Sum*
1. Define $P_1, P_2, \ldots, P_n$ as $n$ parties where $n >= 4$.
2. Assume these parties have secret inputs $x_1, x_2, \ldots, x_n$.
3. Each party $P_i$ breaks its data $x_i$ into $k = n$ segments $d_{i1}, d_{i2}, \ldots, d_{ik}$ where $\sum d_{ij} = x_i$ for $j = 1$ to $k$.
4. Arrange parties in a ring as $P_1, P_2, \ldots, P_n$ and select $P_1$ as the protocol initiator.
5. Assume $rc = k$ and $S_{ij} = 0$. /* $S_{ij}$ is partial sum and $rc$ is round counter*/
6. While $rc != 0$
   begin
   for $j = 1$ to $k$
    begin
    for $i = 1$ to $n$
     begin
       starting from $P_1$ each party computes cumulative sum $S_{ij}$ of its next segment and the received sum from its neighbor and sends to the next party in the ring
     end
    $P_1$ exchanges its position with $P_{(j+1) \mod n}$
    end
    $rc = rc - 1$
   end
7. Party $P_1$ announces the result as $S_{ij}$.
8. End of algorithm.

### 3.2.3 Performance Analysis of the *Modified ck-Secure Sum* Protocol

In *Modified ck-Secure Sum* Protocol the parties pass their data segments to the next party in the ring. It maintains the privacy of the actual inputs. The sum of these segments is same as the sum of the original inputs. Now, if two colluding neighbors try to know the data of a middle party, they cannot do so because the protocol does not allow any two parties to be neighbors of a party for all the rounds of the computation. The neighbors are changed at least once during the computation. Now, if more than two (less than $n$-1) parties want to know the data of a party, they will also be unable to do so because at least one segment will be unknown to them. Thus, only $n$-1 parties out of $n$ can hack the data of a single victim. This is an appreciable improvement over the *ck-Secure Sum* Protocol which guaratees privacy only against two colluding neighbors.

Number of rounds of computation is $n$ and the number of exchanges between parties is $n$–1. The only limitation of this scheme is that the topology of the computational network changes in each round of the computation.

The communication and computation complexity both come to be $nk$ because in one round of the computation there would be $n$ computations and $n$ communications. Thus we can write communication complexity $C(n)$ and computation complexity $S(n)$ as shown in (1) and (2).

$C(n) = n.k = n (n)$

$$C(n) = n^2 \qquad (1)$$

$$S(n) = n^2 \qquad (2)$$

The communication and computation complexity both are $O(n^2)$. Figures 4 depicts the communication and the computation complexity for *Modified ck-Secure Sum* protocol with the number of parties.

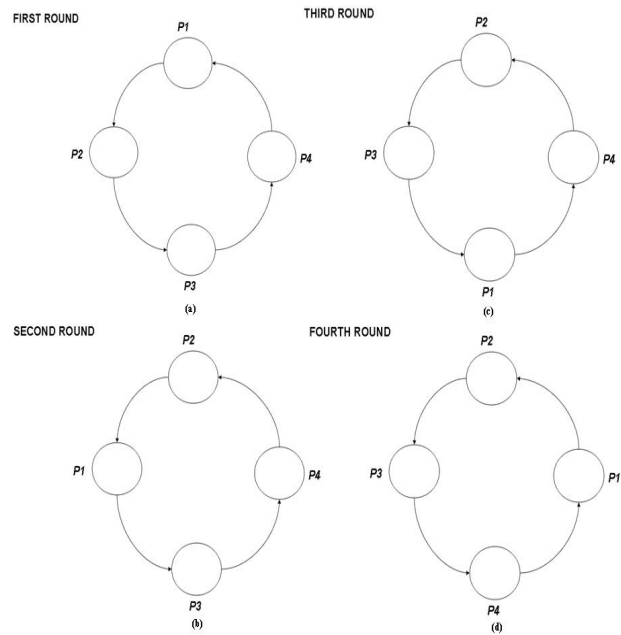

Fig. 2. Snapshots of *Modified ck-Secure Sum* Protocol for four-party case.

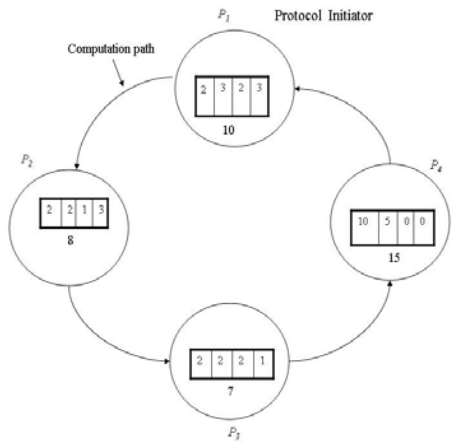

Fig. 3. Snapshots of *Modified ck-Secure Sum* Protocol for four-party case.

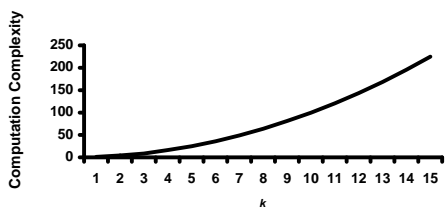

Fig. 4. Computation and Computation Complexity as a function of no. of parties

## 4 CONCLUSION AND FUTURE SCOPE

Secure sum computation is an important component of toolkit for SMC. The secure sum protocols are needed for secure sum computation with lower probability of data leakage. In this paper, the protocol provides zero probability of data leakage by two or more colluding parties which want to know the data of some party. In this protocol, the data block of each party is broken into certain number of segments and computation is performed over these segments. The parties are allowed to change their position in the ring. This ensures that a party can not have same neighbors for all the rounds of the computation. Thus two or more colluding parties cannot learn the secret data of some other party. This is an improvement over previous protocols which ensure safety for two colluding neighbors only. Further efforts can be done to design and analyze the protocol for malicious parties who neither follow the protocol nor honor the privacy of the parties. Protocols can be designed to make the data secure in case majority of the parties are semi honest.

## Authors Profile

*Rashid Sheikh*

M.Tech. Student, Shri Satya Sai Institiute of Science and Technology,
Sehore, M. P., India,
Ph. +91 9826024087
Email: rashidsheikhmrsc@yahoo.com

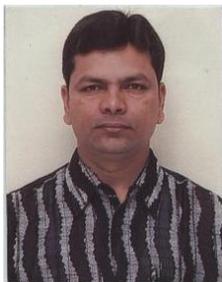

**Biography**: Rashid sheikh has received B.E.(Bachelor of Engineering ) degree in Electronics and Telecommunication Engineering from Shri Govindram Seksaria Institute of Technology and Science, Indore, M.P., India in 1994. He has 15 years of teaching experience. His subjects of interest include Computer Architecture, Computer Networking, Electrical Circuit analysis, Digital Computer Electronics, Operating Systems and Assembly Language Programming. Presently he is pursuing research work for his M. Tech. (Computer Science and Engineering) thesis at SSSIST, Sehore, M.P., India. He has published several research papers in National/International Conferences and Journals. His research areas are Secure Multiparty Computation and Mobile Ad hoc Networks. He is the author of ten books written on Computer Organization and Architecture.

*Beerendra Kumar*
Ph. +91 9770435336
Email: beerucsit@gmail.com

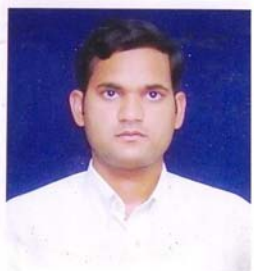

Beerendra Kumar has received B.Tech. (Bachelor of Technology) degree in Computer Science and Information Technology from Institute of Engineering and Technology, Rohilkhand University, Bareilly (U.P), India in 2006. He has completed his M.Tech. (Master of Technology) in Computer Science from SCS, Devi Ahilya University, Indore, India in 2008. He has two years of teaching experience. His subjects of interest include Computer Networking, Theory of Computer Science, Data Mining, Operating Systems and Analysis & Design of Algorithms. He has published three research papers in national conferences and one research paper in international journal. His research areas are Computer Networks, Data Mining, Secure Multiparty Computations and Neural Networks.

*Dr. Durgesh Kumar Mishra*

Professor (CSE) and Dean (R&D),
Acropolis Institute of Technology and Research, Indore, MP, India,
Ph - +91 9826047547, +91-731-4730038
Email: durgeshmishra@ieee.org

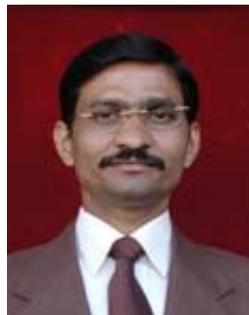

**Biography:** Dr. Durgesh Kumar Mishra has received M.Tech. degree in Computer Science from DAVV, Indore in 1994 and PhD degree in Computer Engineering in 2008. Presently he is working as Professor (CSE) and Dean (R&D) in Acropolis Institute of Technology and Research, Indore, MP, India. He is having around 20 Yrs of teaching experience and more than 5 Yrs of research experience. He has completed his research work with Dr. M. Chandwani, Director, IET-DAVV Indore, MP, India in Secure Multi- Party Computation. He has published more than 60 papers in refereed International/National Journal and Conference including IEEE, ACM etc. He is a senior member of IEEE with Chairman IEEE Computer Society Bombay Section and Vice Chairman IEEE MP-Subsection, India. Dr. Mishra has delivered his tutorials in IEEE International conferences in India as well as other countries also. He is also the programme committee member of several International conferences. He visited and delivered his invited talk in Taiwan, Bangladesh, USA, UK etc in Secure Multi-Party Computation of Information Security. He is an author of one book also. He is also the reviewer of tree International Journal of Information Security. He is a Chief Editor of Journal of Technology and Engineering Sciences. He has been a consultant to industries and Government organization like Sale tax and Labor Department of Government of Madhya Pradesh, India.